\documentstyle[12pt,epsf]{article}

\newcommand{\be}{\begin{equation}}
\newcommand{\ee}{\end{equation}}
\newcommand{\bea}{\begin{eqnarray}}
\newcommand{\eea}{\end{eqnarray}}
\newcommand{\beano}{\begin{eqnarrayno}}
\newcommand{\eeano}{\end{eqnarrayno}}
\newcommand{\nnm}{\nonumber}

\def\df{\ifnextchar[{\df@a}{\df@b}}
\def\df@a[#1]#2{{\rm d}^{#1}#2\,}
\def\df@b#1{{\rm d}#1\,}

\newcommand{\ket}[1]{\mbox{$|#1\rangle$}}
\newcommand{\braket}[2]{\mbox{$\langle #1 | #2 \rangle$}}
\newcommand{\brakket}[3]{\mbox{$\langle #1 | #2 | #3 \rangle$}}

\newcommand{\mexp}[1]{\langle m_{#1} \rangle}

\title{A microscopic semiclassical confining field equation for $U(1)$
lattice gauge theory in 2+1 dimensions}
\author{Christoph Best,\thanks{email: {\tt cbest@th.physik.uni-frankfurt.de}}
        \ Andreas Sch\"afer \\
        {\small Institut f\"ur Theoretische Physik, J. W.
Goethe-Universit\"at,}\\
        {\small 60054 Frankfurt am Main, Germany}
}
\date{January 23, 1996}

\begin{document}

\maketitle

\begin{abstract}
We present a semiclassical nonlinear field equation for the confining field
in 2+1--dimensional $U(1)$ lattice gauge theory (compact QED). The equation
is derived directly from the underlying microscopic quantum Hamiltonian by
means of truncation. Its nonlinearities express the dynamic creation of
magnetic monopole currents leading to the confinement of the electric field
between two static electric charges. We solve the equation numerically and
show that it can be interpreted as a London relation in a dual
superconductor.
{\em (UFTP preprint 405/1996)}
\end{abstract}

\section{Introduction}

In the last few years, several studies of confinement in lattice gauge
theories have succeded in observing
not only the confinement potential \cite{Bali1,Booth}, but also
the formation of flux tubes between two
static charges \cite{Bali,Haymaker}.
The most promising mechanism for this effect is the dual
superconductor hypothesis \cite{thooft,Mandelstam,BBZ} which assumes that the
electric field is confined by magnetic monopole currents forming the walls
of the flux tube, just like superconducting electric currents confine the
magnetic field in an ordinary superconductor. The monopoles are presumed to
originate dynamically from the periodicity of the Hamiltonian as tunneling
effects between neighboring minima \cite{Banks,Stack}. Abelian monopoles
occur naturally in $U(1)$ lattice gauge theory (compact QED). As $U(1)$ is a
subgroup of $SU(2)$ and $SU(3)$, so-called ``Abelian'' monopoles (monopoles
in the maximally Abelian projection) \cite{tHooft2}
may also play a role in the QCD
mechanism of confinement. The dynamic creation of magnetic monopoles and
their description by a London relation has been observed in several studies
\cite{DGT,Singh,DiG,Cea,DelDebbio} based on DeGrand and Toussaint's
identification prescription \cite {DGT} for monopoles.

Recently, Zach et.~al.~\cite{Zach} demonstrated that confinement of the
electric field in $U(1)$ lattice gauge theory can be effectively described
by the Maxwell-London equations of a dual superconductor. This model has
only a single parameter, the London penetration length. It is, however,
phenomenological in the sense that it is not derived from the underlying
microscopical lattice system.

Most studies have employed Wilson's Euclidean picture of lattice gauge
theory \cite{Wilson}, simulating it as a statistical system using Monte
Carlo methods. While this offers easy access to the confining potential
(string tension) by measuring the expectation value of a Wilson loop, and to
the monopole density by means of the DeGrand-Toussaint prescription,
measurements of the fields involves a more complex correlation function
(basically between a loop and a plaquette) and thus more
statistical fluctuations.

An alternative approach is to make use of the Kogut-Susskind Hamiltonian
formulation \cite{KS}, considering the theory as a quantum system with many
degrees of freedom and applying methods from many-body theory as developed
in e.g.~computational chemistry or nuclear physics
\cite{Morningstar,Lsm,Bishop,Chin,Koonin,CB1}. In particular, the
Hamiltonian formulation provides more easily for a semiclassical static
solution than the Euclidean version.

In the following, we perform an {\em ab initio} derivation of a
semiclassical field equation in $2+1$ dimensions that incorporates the prime
feature of the gauge group $U(1)$---the periodicity of the action that leads
to the dynamic creation of effective monopole currents. The resulting
nonlinear field equation exhibits confinement of the electric field caused
by magnetic monopole currents. It is microscopic in the sense that it is
formulated in terms of the plaquette variables (i.e.~the vortex potential of
the electric field) and does not make use of phenomenologically introduced
monopole fields.

We first review the formulation of classical electrodynamics on the lattice.
After introducing the Kogut-Susskind Hamiltonian, we move to the plaquette
representation (which only exists without constraints in 2+1 dimensions) and
then perform a truncation of the Hilbert space to obtain a governing
equation for plaquette wavefunctions. By studying single-plaquette
solutions, we are led to an effective field equation in terms of the
plaquette variables. Its linearized form is the London relation. The
numerical solution shows confinement of the electric field. We discuss the
consequences of this for the confinement of charges in lattice gauge
theories.

\section{Derivation of the effective field equations}

\subsection{Classical electrodynamics on the lattice}

We are interested in the distribution of the electric field $E_l$ on the
links $l$ of a 2-dimensional rectangular lattice in the presence of static
charges. In the continuum, the electric field is determined by two
conditions
\bea
  \vec\nabla \cdot \vec E(\vec r) = \rho_e(\vec r) \quad ,\\
  \vec\nabla \times \vec E(\vec r) = 0 \quad .
\eea
The first is Gauss' law, expressing that the electric charge density is the
source of the electric field, the second Stokes' law that there are no
closed lines of force (when we consider the static case without currents).
These two equations determine the electric field completely. If the field
exhibits confinement, one or both of them have to be modified. As the first
equation is linked to gauge invariance of the theory (which is not
violated), the only possible modification is to replace the zero on the
right-hand side of the curl equation by a quantity $- J$, interpreted as
the density of magnetic monopole currents.

On the lattice, the electric field is represented by its flux $E_l$ along
the directed links $l$. Gauss' law then becomes a condition on the link
fields $E_l$ joined at a site $s$
\be \label{gauss}
  \sum_l \epsilon_{s,l} E_l = \rho_l
\ee
where $\epsilon_{s,l}$ is $+1$ if the link $l$ leads away from site $s$,
$-1$ if it leads to site $s$, and zero otherwise. In other words, the
difference of inbound and outbound flux at any site equals the charge at
this site.

Stokes' law turns into a condition on each plaquette $p$,
\be \label{stokes}
  \sum_l \epsilon_{p,l} E_l = -J_p
\ee
where $\epsilon_{p,l}$ is $+1$ or $-1$ if the directed link $l$ counts
positively or negatively, resp., when going counter-clockwise around the
plaquette. This condition states that the electric flux around a
plaquette is generated by magnetic monopole currents.

\subsection{Quantum electrodynamics on the lattice}

In quantum field theory, the electromagnetic field is governed by the
Kogut-Susskind Hamiltonian \cite{KS} for the gauge group $U(1)$:
\be\label{KSH}
  H = - \frac{1}{2} \sum_{l\in\mbox{\scriptsize links}}
      \frac{\partial^2}{\partial^2 \Theta_l}
      + \lambda \sum_{p\in\mbox{\scriptsize plaquettes}} \left(
      1 - \mbox{Re} \, e^{i\Theta_p} \right) \quad,
\ee
where $\lambda = 1/g^4$, $\Theta_l$ is the group parameter associated with
the link $l$, and $\Theta_p$ the counter-clockwise sum of the link
parameters around the plaquette $p$,
\be
  \Theta_p = \sum_l \epsilon_{p,l} \Theta_l \quad.
\ee
The Hamiltonian operates on a Hilbert space spanned by basis states
\be
  \ket{\Theta} = \prod_{l\in\mbox{\scriptsize links}}
       \ket{\Theta_l}, \qquad -\pi \le \Theta_l < \pi \quad.
\ee
The first (kinetic) term in eq.~(\ref{KSH}) is associated with the electric
field, the second (potential) term with the magnetic field. Without the
nonlinearity introduced by the higher-order terms of the cosine in the
magnetic field term, this Hamiltonian would describe the motion of
(nonrelativistic) particles in a harmonic potential (noncompact QED).
In compact QED, the harmonic potential is replaced by a periodic potential
that can be effectively described, as we shall see, by the dynamic creation
of monopole currents.

As we are interested in the electric field, it is advantageous to switch to
eigenstates of the momentum operator
\be \label{momop}
   \hat n_l = -i\frac{\partial}{\partial \Theta_l}
\ee
given by
\be
  \braket{\Theta_l}{n_l} = e^{in_l\Theta_l} \quad.
\ee
The Hilbert space will then be spanned by basis states $\ket{n}$ whose
label $n$ stands for a set of integer numbers $n_l$, one for each link $l$:
\be
  \ket{n} = \prod_{l\in\mbox{\scriptsize links}}
            \ket{n_l}, \quad n_l \in {\bf Z} \quad.
\ee
In this representation, the Hamiltonian can be written
\be \label{Ham}
  H = \frac{1}{2} \sum_{l\in\mbox{\scriptsize links}}
      \hat n^2_l
      + \lambda \sum_{p\in\mbox{\scriptsize plaquettes}} \left(
      1 - \frac{1}{2} \hat U_p - \frac{1}{2} \hat U^+_p \right) \quad,
\ee
where $n$ represents the momentum operator (\ref{momop}) with the matrix
element
\be
  \brakket{n'}{\hat n}{n} = n \, \delta_{n,n'} \quad,
\ee
and the group element operator $\hat U_p$
\be
  \hat U_p = e^{i\Theta_p}
\ee
with the matrix elements
\be
  \brakket{n'}{\hat U_p}{n} = \delta_{n',n+p} \quad, \qquad
  \brakket{n'}{\hat U^+_p}{n} = \delta_{n',n-p} \quad,
\ee
where $n'-p$ represents the configuration $n'$ with all links in the
plaquette $p$ raised or lowered by one unit according to whether they count
positively or negatively when traversing the links of the plaquette
counter-clockwise.

Note that the periodicity of the action has now transformed into the
discretization of the dynamic variable $n$. This is the only nonlinearity
in this equation.

The Hamiltonian ($\ref{Ham}$) commutes with the operators
\be
  \hat G_s = \sum_{l\in\mbox{\scriptsize links}}
        \epsilon_{s,l} \hat n_l, \qquad s\in\mbox{sites}
\ee
that generate gauge transformations at the lattice sites $s$. In the
electric field representation, these operators are diagonal and their
eigenvalues give the total electric field flux out of the sites $s$.
By means of Gauss' law (\ref{gauss}), they are associated with static
charges sitting at the lattice sites $s$. The eigenstates of the Hamiltonian
separate into different sectors, each associated with a certain
distribution of integer charges over the lattice sites. We are especially
interested in the string sector with two opposite charges a certain
distance away from each other.

In the strong-coupling limit $\lambda \to 0$, the Hamiltonian is dominated
by the electric term. The lowest eigenstate in the string sector is the
shortest path of raised links connecting the two charges. As the magnetic
contribution is switched on by raising $\lambda$, perturbations are created
from the original ground state by the action of the plaquette raising and
lowering operators $U_p^+$ and $U_p$.

\subsection{Plaquette representation}

In two dimensions, self-duality of the lattice allows us to introduce new
variables $m_p$ associated with plaquettes $p$ to replace the link variables
$n_l$ by the relation
\be \label{defM}
  n_l = \sum_{p} \epsilon_{p,l} m_p + N_l
\ee
As there are exactly two plaquettes $p$ for which $\epsilon_{p,l}$ is
nonzero, namely the plaquettes to the left ($\epsilon_{p,l}=1$) and to the
right ($\epsilon_{p,l}=-1$) of the link $l$, (\ref{defM}) expresses each
link as the difference of the two adjacent plaquettes. The corresponding
continuum relation is
\be
  \vec E = \nabla \times \vec m + \vec N
\ee
where the vector field $\vec m$ is in the $z$-direction such that $\vec E$
is in the $x$-$y$-plane. Thus $\vec m$ is the vortex potential of the
electric field. Gauss' law is automatically satisfied
\be
  \vec\nabla \cdot \vec E = \vec\nabla \cdot \vec N = \rho_e
\ee
given an appropriate choice of $\vec N$ as a representative for the sector
of the theory, e.g.~a Dirac string for the two-charge sector. There is a
residual gauge invariance under global shifts of the $m_p$.

The original Hilbert space is represented by the direct product of the set
of all plaquette variable configurations $m_p$ and one configuration $N_l$
representative for each sector of the theory. It thus allows us to focus our
attention on one sector by fixing $N_l$ while admitting all configurations
$m_p$.

Turning first to the classical equations (\ref{gauss}) and (\ref{stokes}),
we note that, while Gauss' law is automatically satisfied, Stokes' law
becomes
\be \label{pstokes}
  4 m_p - \sum_{(pp')} m_{p'} + N_p \equiv - \Delta_L m_p + N_p = - J_p
\ee
where the sum runs over all neighbors $p'$ of the plaquette $p$, $\Delta_L$
is the lattice Laplace operator on an unit lattice, and $N_p$ is defined as
the lattice curl of $N_l$
\be
  N_p = \sum_l \epsilon_{p,l} N_l \quad.
\ee
In particular, if $N_l$ is a string connecting two static charges, $N_p$ is
$+1$ on the plaquettes just above the string and $-1$ just below the string.

The quantum Hamiltonian in the plaquette representation reads
\be \label{plaqHam}
  H = 2 \sum_p \hat m_p^2 - \sum_{(pp')} \hat m_p \hat m_{p'}
     + \frac{1}{2} \sum_l N_l^2
     + \sum_p \hat m_p N_p
     + \lambda \sum_p \left( 1 - \frac{1}{2} \hat U_p
                               - \frac{1}{2} \hat U^+_p \right)
\ee
where $(pp')$ sums over neighboring plaquettes $p'$. We note that, while the
diagonal (potential) part of (\ref{plaqHam}) is quadratic and thus amenable
to a Gaussian' solution, the operators $U_p$ are highly nonlinear as they
implicitly contain higher powers of the derivative operator.
This, of course, is an expression of the fact that $m$ is confined to
integer values which, in turn, results from the periodicity of the original
$U(1)$ action and will ultimately give rise to a modification of Stokes'
equation. We also note that this restriction is enforced by the definition
of $\hat U_p$ alone so that we may consider $m$ a continuous variable.

\subsection{Truncation}

To solve the eigenvalue problem of eq. ($\ref{plaqHam}$), we truncate the
eigenvalue equation to states that are direct products of single-plaquette
states,
\be
  \ket{\Psi} = \prod_{p} \ket{\psi_p} \quad.
\ee
To find the governing equation for plaquette $q$, we project onto the state
\be
  \ket{\Psi'_q} = \prod_{p\ne q} \ket{\psi_p} \quad.
\ee
This leads to the relation
\bea
  \brakket{\Psi'_q}{H}{\Psi} &=&
  \left( \sum_{p\ne q} \brakket{\psi_p}{H_p}{\psi_p}
        +\sum_{p\ne q} \sum_{p'\ne q}
         \brakket{\psi_p \psi_{p'}}{H_{pp'}}{\psi_p \psi_{p'}} \right)
  \, \ket{\psi_q} \nnm\\
  &+& H_q \ket{\psi_q}
     +\sum_{p\ne q} \brakket{\psi_p}{H_{qp}+H_{pq}}{\psi_p \psi_q} \nnm\\
  &=& E \ket{\psi_q}
\eea
where the Hamiltonian was decomposed according to
\be
  H = \sum_p H_p + \sum_{pp', p\ne p'} H_{pp'} + H_0
\ee
($H_0$ accounts for the constant part in the Hamiltonian).
In this approximation, we find as the governing equation for the plaquette
wavefunction $\ket{\psi_p}$
\be \label{plaqHam2}
  \left[  2 \hat m_p^2
         - \sum_{(pp')} \hat m_p \mexp{p'}
     + \hat m_p N_p
     + \lambda \left( 1 - \frac{1}{2} \hat U_p - \frac{1}{2} \hat U^+_p
\right)
  \right] \ket{\psi_p}
  = E_p
    \ket{\psi_p} \quad,
\ee
that is, a single-plaquette equation with the next-neighbor interaction
replaced by a mean-field term. The single-plaquette energy $E_p$ is
related to the total energy by
\be \label{energ}
  E_p = E
        - \sum_{q\ne p} \langle H_q \rangle
        - \sum_{q\ne p} \sum_{q'\ne p} \langle H_{qq'}\rangle - H_0 \quad.
\ee
Summing over all $E_p$ would double-count the next-neighbor interactions.
The correct ground-state energy is given by the expectation value of the
Hamiltonian. However, since we have dropped all correlations between
plaquettes, its value in this approximation cannot be trusted.

This approach can be extended to basis functions defined on a blocked
lattice, as would be required for a renormalization group analysis.

\subsection{Single-plaquette solutions}

We rewrite the single-plaquette equation
\be
  H_{\rm s.p.} = 2 \hat m^2 - J \hat m +
                 \lambda \left( 1 - \frac{1}{2} \hat U
                                  - \frac{1}{2} \hat U^+ \right)
\ee
introducing a generic source term $-J\hat m$ that accounts for both the
original source term and for the next-neighbor interaction:
\be \label{Jeq}
  J_p = \sum_{(pp')} \mexp{p'} - N_p
  = \Delta_L m_p + 4 m_p - N_p
  \quad.
\ee
We shall discuss this equation by considering its solution in the strong-
and the weak-coupling limit.

In the strong-coupling limit $\lambda\to 0$, the Hamiltonian reduces to
\be
  2 \hat m^2 - J \hat m \quad.
\ee
As it is diagonal in $m$, eigenfunctions are concentrated on a single value
of $m$:
\be
  \braket{m}{\psi_n} = \delta_{n,m} \quad.
\ee
The state associated to the lowest eigenvalue is found by minimizing the
energy over the integer numbers:
\bea
  E_{\rm min} &=& \mbox{min}_{m \in {\bf Z}} (2 m^2 - Jm)
                 = - \frac{J_0^2}{8} + \frac{(J-J_0) J_0}{4} \nnm\\
  m_{\rm min} &=&
  \left[ \frac{J}{4} + \frac{1}{2} \right]
  \equiv \frac{J_0}{4}
\eea
where $[x]$ denotes the smallest integer not larger than $x$ and thus $J_0$
the multiple of $4$ closest to $J$.

In the weak-coupling limit $\lambda \to \infty$, we assume that the
ground-state wave function $\braket{m}{\psi}$ is spread out sufficiently
wide that it can be approximated by a Gaussian:
\be
  \braket{m}{\psi} \approx N \exp \left( - \frac{(m-m_0)^2}{4\sigma} \right)
\ee
with a normalization constant $N$, first moment $m_0$ and second moment
$\sigma$. We have
\be
  \brakket{m}{U}{\psi} = \braket{m+1}{\psi}
  = e^{-(1 + 2(m-m_0))/4\sigma} \braket{m}{\psi}
\ee
and thus
\bea
  \brakket{m}{H}{\psi}
  &=& 2m^2 - Jm  \nnm\\
  &+& \lambda \left[ 1 - \frac{1}{2} e^{-1/4\sigma}
                    \left( e^{(m-m_0)/2\sigma} + e^{-(m-m_0)/2\sigma} \right)
            \right] \braket{m}{\psi} \nnm\\
  &=& E \braket{m}{\psi} \quad.
\eea
Assuming that $\sigma \gg 1$ and considering values of $m$ in the vicinity
of $m_0$, we can assume $|(m-m_0)/\sigma| \ll 1$ and expand the
exponentials. Keeping terms up to $1/\sigma^2$, we find the eigenvalue
equation
\be
  2 m^2 - Jm + \lambda \left[
      1 + \frac{1}{8\sigma}
        - \left( 1 + \frac{(m-m_0)^2}{8\sigma^2} \right)
  \right]
  = E \quad.
\ee
To satisfy this equation for any $m$, we set
\be
  \sigma = \frac{\sqrt{\lambda}}{4} \quad, \qquad
  m_0 = \frac{J}{4} \quad.
\ee
The eigenvalue is
\be
  E = \sqrt{\lambda} - \frac{J^2}{8} \quad.
\ee

\begin{figure}[htb]
\centerline{\epsfbox{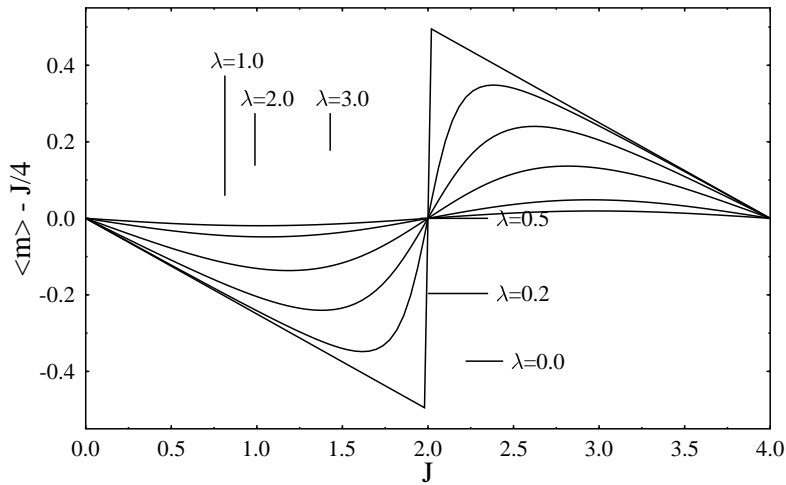}}
\caption{Deviation of the expectation value of the plaquette variable
$\mexp{}$
from its weak-coupling limit as a function of the external source $J$
at different couplings $\lambda$. This quantity is a measure for the
nonlinearity of the system.
\label{fig1}}
\end{figure}

We thus find both in the strong- and weak-coupling limit that the
expectation value of the plaquette variable $m$ is approximately
\be
  \brakket{\psi}{m}{\psi} \equiv \mexp{} \approx \frac{J}{4} \quad.
\ee
However, while this becomes exact in the weak-coupling limit, in strong
coupling the wave function feels the restriction of $m$ to integers.
Fig.~\ref{fig1} shows the deviation of the numerically calculated
expectation value of $m$ from the weak-coupling expression as $J$ is varied.
It exhibits a sawtooth shape in strong-coupling that is softened to a
sine-like shape in the intermediate range. The cause of the deviation is
the restriction of $m$ to integer values which is a direct consequence of
the periodicity of the action.

We can thus parametrize the nonlinearity of the single-plaquette Hamiltonian
by writing
\be \label{meq}
  \mexp{\rm approx.} = \frac{J}{4} - \alpha \sin \left( \frac{\pi}{2} J
\right) \quad.
\ee
The parameter $\alpha$ is a function of $\lambda$ and vanishes in the
weak-coupling limit. In the strong-coupling limit $\lambda\to 0$, we find
instead
\be \label{meqsc}
  \mexp{\rm s.c.} = \frac{J}{4} - \frac{J - J_0}{4}
\ee
where $J_0$ again is the closest multiple of 4 to $J$.

\subsection{Effective field equation}

An effective field equation can be obtained by substituting eq.~(\ref{Jeq})
into eq.~(\ref{meq}), leading to
\be
  m_p = \frac{1}{4} \sum_{pp'} m_{p'} - \frac{1}{4} N_p
          - \alpha \sin \frac{\pi}{2} \left( \sum_{pp'} m_{p'} + N_p
                                      \right)
\ee
where we dropped the expectation value brackets for $m_p$.
This is equivalent to
\be \label{theeqn}
  - \Delta_L m_p + N_p = - J_p =
  - 4\alpha \sin \left[ \frac{\pi}{2}
    \left( \Delta_L m_p + 4 m_p - N_p \right)\right] \quad.
\ee
We recognize on the left-hand side Stokes' law (\ref{pstokes}).
Consequently, the new term on the right-hand side represents $J_p$, the
density of monopole currents. It depends nonlinearly on itself and on the
plaquette variable $m_p$. In actual calculations, we use the numerical
result from Fig.~\ref{fig1} in place of the sine. The periodicity of the
right-hand side with respect to $m_p$ assures the residual gauge invariance
under a global integer shift.

To get a first impression of the meaning of eq.~(\ref{theeqn}), we
linearize the sine. This gives after some algebra
\be
  \left( -\Delta_L + \frac{8\pi\alpha}{1 - 2\pi\alpha} \right)
  \, m_p + N_p = 0 \quad.
\ee
Using Stokes' law (\ref{pstokes}), this is in continuum notation
\be
  - \vec J + \frac{8\pi\alpha}{1-2\pi\alpha} \vec m = 0 \quad.
\ee
By taking the curl, one obtains
\be \label{EN}
  \vec E = \frac{1-2\pi\alpha}{8\pi\alpha} \, \vec\nabla \times \vec J_m
           + \vec N
\ee
which is a London relation between the electric field and the monopole
current (cf.~eq.~(4) and (5) of \cite{Singh} and eq.~(15) of \cite{Zach};
note also that the Dirac string on the right-hand side comes in naturally in
our formulation). The London penetration length is in this approximation
\be
  \sqrt{\frac{1-2\pi\alpha}{8\pi\alpha}} \quad.
\ee
It duly diverges as $\alpha\to 0$ (weak-coupling limit). At a critical value
of
\be
  \alpha_{\rm crit} = \frac{1}{2\pi}
\ee
the London penetration length becomes zero and then imaginary, thus
signalling the breakdown of the linearizing approximation. At this point,
eq.~(\ref{EN}) reduces to
\be
  \vec E = \vec N \quad,
\ee
i.e.~the electric field is the Dirac string.

In the strong-coupling limit, we can use without the linearization
eq.~(\ref{meqsc}) for the right-hand side. This leads to
\be
  -\Delta_L m_p + N_p = - (\Delta_L m_p + 4 m_p - N_p)
                       + 4 \left[ \frac{\Delta_L - N_p + m_p}{4}
                       + \frac{1}{2} \right]
\ee
or equivalently
\be
  m_p = \left[ \frac{\Delta_L m_p - N_p + 4m_p}{4} + \frac{1}{2} \right ]
\quad.
\ee
The brackets again indicate rounding down to an integer. This equation is
in particular satisfied for $m_p \equiv 0$, which again corresponds to
\be
  \vec E = \vec N \quad,
\ee
i.e.~the Dirac string.

\section{Numerical solution and Discussion}

The nonlinear partial differential equation (\ref{theeqn})
can be solved numerically in certain circumstances. We used a simple
iterative method in which the equation is solved by simultaneous
over-relaxation for a fixed right-hand side. The right-hand side is then
adjusted to the new values of $\mexp{p}$, and the process repeated until
convergence. We found that this method converges provided that the lateral
extension of the lattice (i.e.~the $y$ dimension) is not too large;
otherwise spontaneous oscillations set in. In particular, the difficult
part here is the strong-coupling region that is highly nonlinear.

\begin{figure}[p]
\centerline{\epsfbox{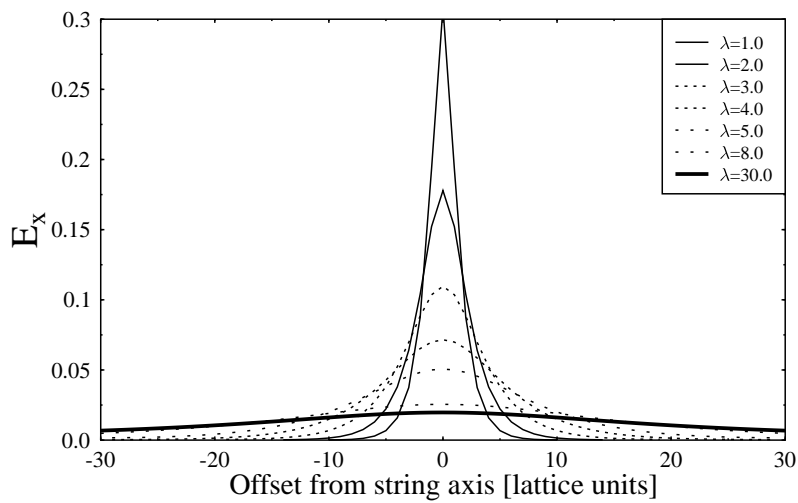}}
\caption{Electric field $E_x$ across the string axis for different
values of $\lambda$ \label{fig2}}
\end{figure}

\begin{figure}[p]
\centerline{\epsfbox{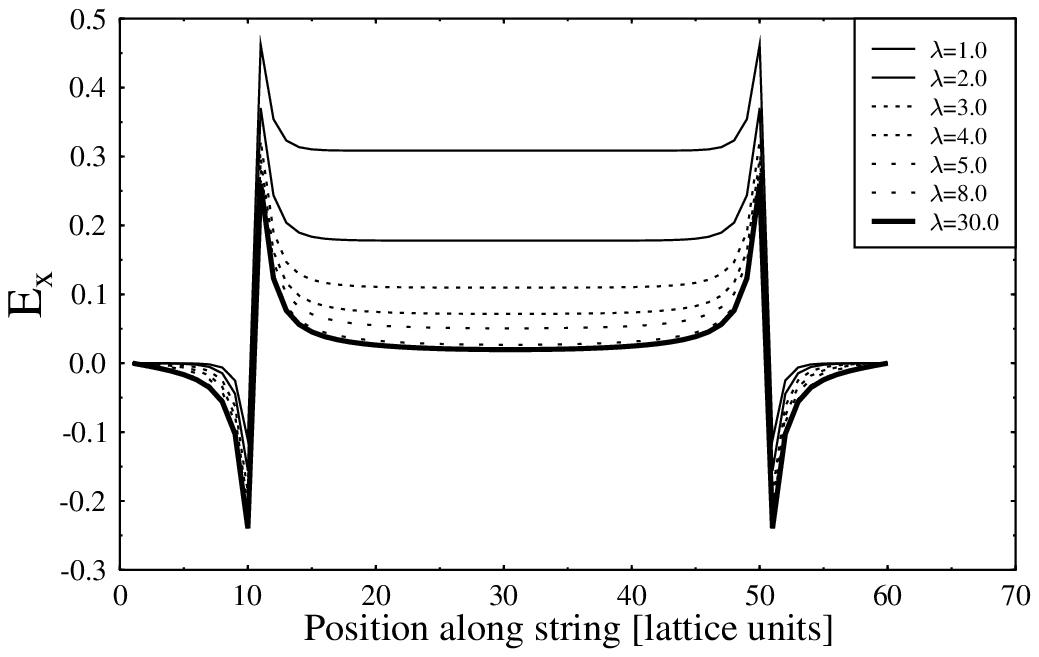}}
\caption{Electric field $E_x$ along the string axis for different
values of $\lambda$ \label{fig3}}
\end{figure}

\begin{figure}[p]
\centerline{\epsfbox{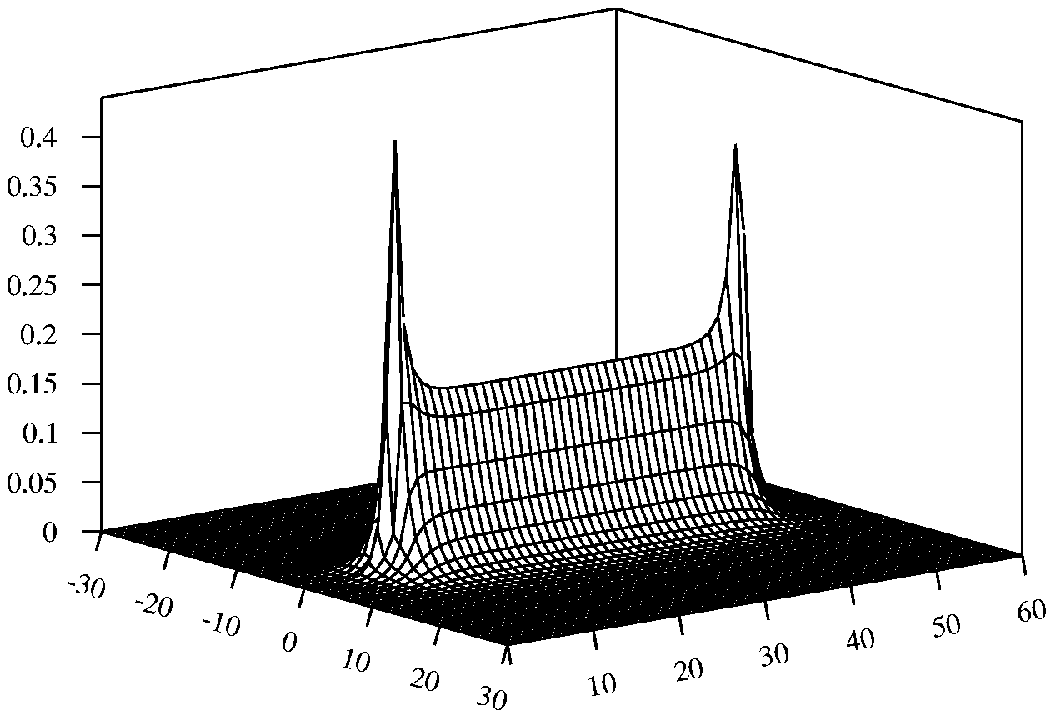}}
\caption{Electrical field strength $|\vec E|$ at $\lambda = 2$ \label{fig7}}
\end{figure}

\begin{figure}[p]
\centerline{\epsfbox{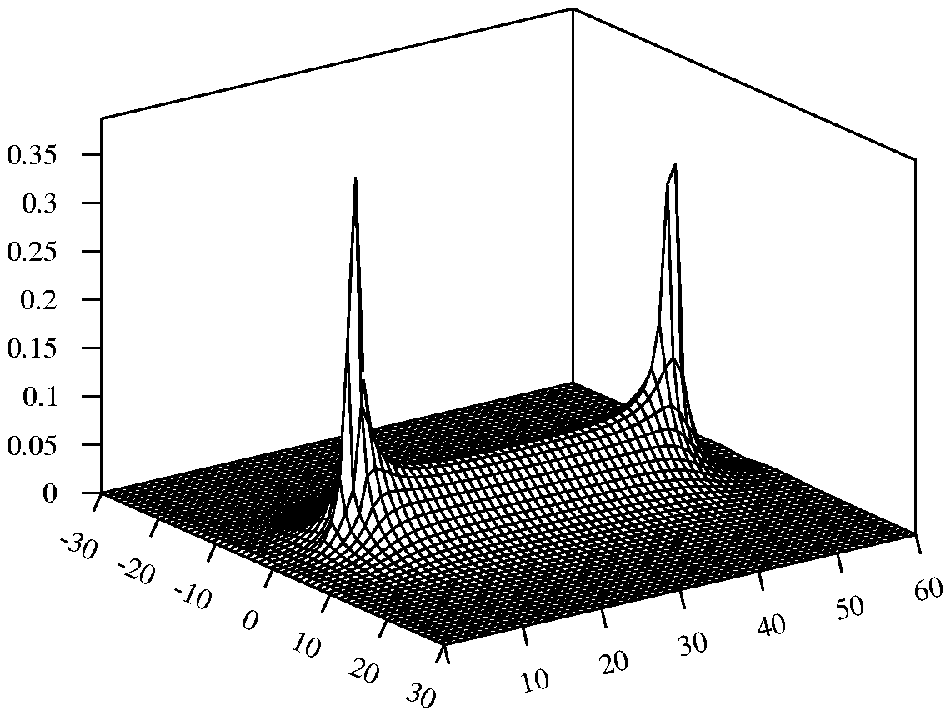}}
\caption{Electrical field strength $|\vec E|$ at $\lambda = 4$ \label{fig9}}
\end{figure}

\begin{figure}[p]
\centerline{\epsfbox{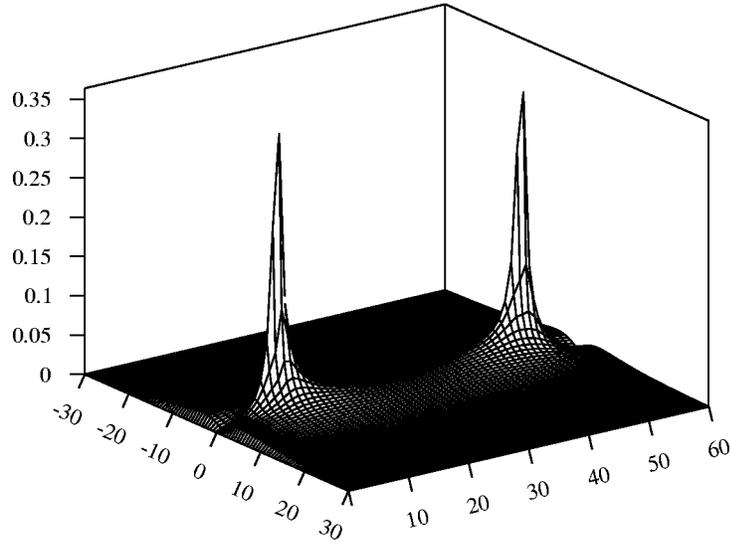}}
\caption{Electrical field strength $|\vec E|$ at $\lambda = 8$ \label{fig8}}
\end{figure}

Fig.~\ref{fig2}--\ref{fig8} show the electric field calculated from
eq.~(\ref{theeqn}) for different values of $\lambda$ at a constant string
length of 40 lattice units. The lattice size was between $60 \times 100$
at weak coupling and $60 \times 22$ at strongest coupling. The solution
varies between the string-like strong-coupling limit and the Coulomb-like
weak-coupling limit.

\begin{figure}[p]
\centerline{\epsfbox{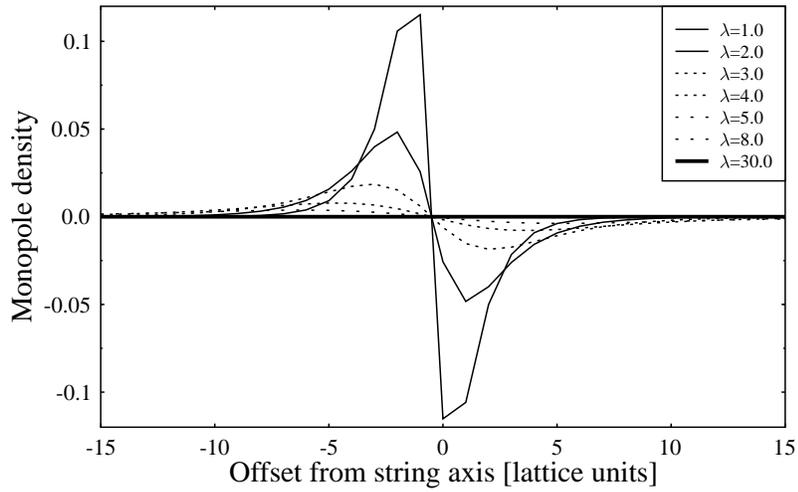}}
\caption{Monopole density in a cut across the string axis for different
$\lambda$\label{fig4}}
\end{figure}

The actual density of monopole charges in a cut through the string is shown
in fig.~\ref{fig4}. Their concentration is largest just next to the string
axis where the field is confined, and then drops sharply. This justifies the
interpretation that the monopole charges create a ``wall'' that confines
the electric field.

\begin{figure}[p]
\centerline{\epsfbox{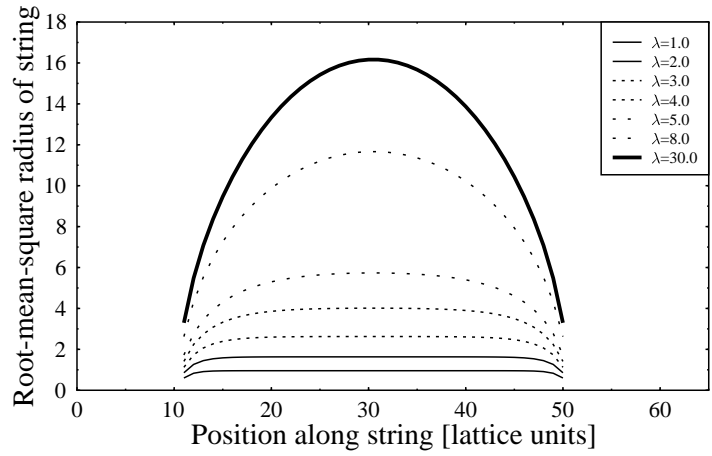}}
\caption{Root-mean-square radius of the string along the string axis for
different values of $\lambda$\label{fig5}}
\end{figure}

\begin{figure}[p]
\centerline{\epsfbox{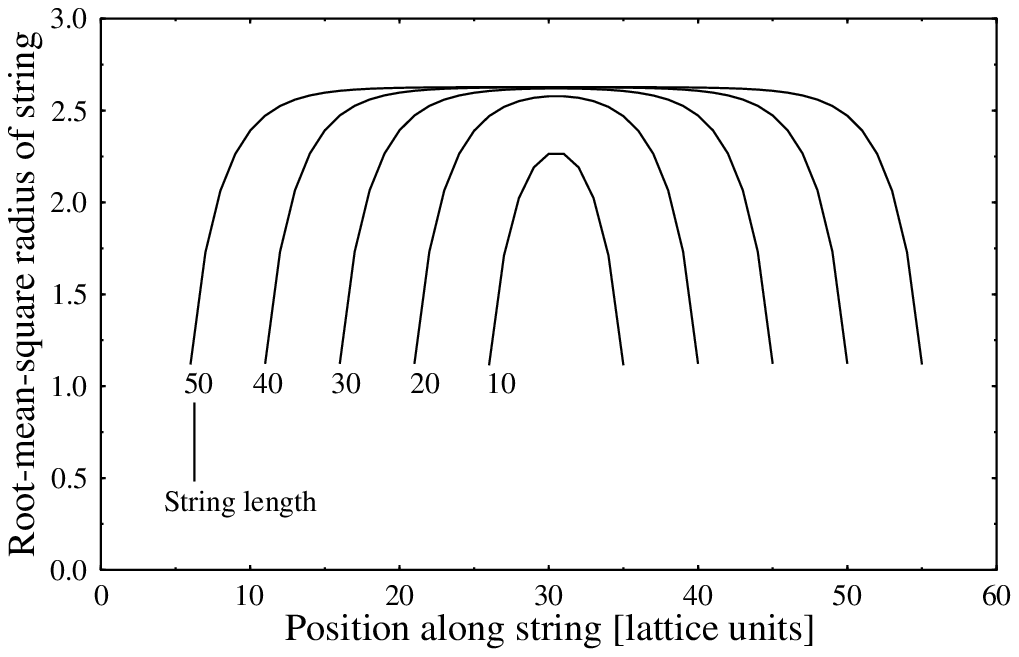}}
\caption{Root-mean-square radius of the string along the string axis for
different distances of the charges\label{fig6}}
\end{figure}

In fig.~\ref{fig5}, we have calculated the root-mean-square radius
\be
  \bar r(x) = \sqrt{\frac{\sum_y y^2 \, E_x^2(x,y)}{\sum_y E_x^2(x,y)}}
\ee
(where $y$ is the distance from the string axis)
orthogonal to the string axis of the flux tube as a function of $\lambda$ and
of the position along the string axis. For large $\lambda$, it is constant
along most of the string axis, indicating the formation of a flux tube. For
small $\lambda$, it smoothly reaches a maximum halfway between the charges
as a result of the $1/r$-law.

Fig.~\ref{fig6} shows the root-mean-square radius for different lengths of
the string. We do not see the $\ln r$ dependence \cite{LMW} predicted from
studying quantum fluctuations of the string. The width of the string stays
constant, as would be expected from strong-coupling expansions. This is an
indication that quantum effects will modify eq.~(\ref{theeqn}).

The numerical solution of eq.~(\ref{theeqn}) exhibits flux tube formation as
is expected for confinement. As $\lambda$ is increased and the weak-coupling
limit is approached, the flux tube disappears.
On the other hand, it is expected from studies of the effective monopole
action
that $2+1$-dimensional $U(1)$ lattice gauge theory should not exhibit a
phase transition to deconfinement \cite{Polyakov}. Both results can be
reconciled when considering that confinement of the electric field does not
at first hand prove confinement of charges or provide for the calculation of
the string tension. For the latter, it has to be shown that the ground-state
energy of the Hamiltonian grows proportional to the distance of the charges.
However, when truncating to single-plaquette states, any information about
the correlation energy is lost, and the ground-state energy cannot be
calculated reliably (cf.~eq.~(\ref{energ})). For such a calculation, blocked
ground states as used in real-space renormalization must be utilized, and
the renormalization group, which is supposed to be the prime cause for
confinement of charges, is brought back in. Quantum effects manifested by a
roughening transition might then modify eq.~(\ref{theeqn}) to reinforce
confinement of the electric field.

\section{Conclusions}

We have shown that a semiclassical confining field equation can be directly
derived from the Hamiltonian formulation of lattice gauge theory. Its
linearized form is a London relation between the electric field and the
monopole current. Numerical calculations show the formation of a flux tube
for the electric field between two charges. The wall of the flux tube is
formed by effective monopole currents that result from the nonlinearities
of the field equation.

These results have been obtained using a rather crude discorrelated vacuum
state. One of the major questions that accompanies this approach is how this
equation is modified by correlations, i.e.~quantum effects, and how this is
related to a possible roughening transition \cite{LMW}. The validity of the
London relation for strong coupling has been demonstrated in Monte Carlo
calculations. For weak coupling, renormalization group analysis predicts
continued confinement of charges while the semiclassical field equations
converge to the limit of classical electrodynamics. Here the inclusion of
correlations (and consequently the renormalization group behavior) is needed
to calculate the string tension and show confinement of charges. The most
promising way to do this is to extend the truncation scheme to correlated
blocks and perform a Hamiltonian real-space renormalization group analysis.

A direct check of the validity of eq.~(\ref{theeqn}) is possible by
comparing the results to Monte Carlo simulations. In particular, we are
investigating the guided random-walk ground-state ensemble projector method
\cite{CB1} that is based on the Hamiltonian approach.

Further extension of the model is to $3+1$ dimensions and to the $SU(2)$
gauge group. In the extension to $3+1$ dimensions, one encounters
constraints in the plaquette representation (Bianchi identities). While this
will probably affect correlations, it remains to be seen whether the
effective field equations changes qualitatively, and how this can be related
to the maximally Abelian gauge which has been used in some of the Monte Carlo
calculations.

Finally the extension to finite temperature can be considered to study the
deconfinement transition.

This work was supported by Deutsche Forschungsgemeinschaft (DFG).
C.~B.~wishes to thank the German National Scholarship Foundation for its
support.

\end{document}